\documentclass[amsmath, amsthm, amssymb, floatfix, reprint, onecolumn, notitlepage, nofootinbib, 12pt]{revtex4-1}
\usepackage{setspace}
\usepackage{amsmath}
\usepackage{breqn}
\usepackage{graphicx}
\usepackage[nearskip,margin = 0pt,caption=false]{subfig}

\usepackage{verbatim}
\usepackage{amsfonts}
\usepackage{amssymb}
\usepackage{epstopdf}
\usepackage{lipsum}
\usepackage{siunitx}
\usepackage{xcolor}
\usepackage{array}
\usepackage{braket}
\usepackage[normalem]{ulem}
\usepackage[resetlabels,labeled]{multibib}
\DeclareGraphicsExtensions{.pdf,.eps,.png,.jpg,.mps}
\usepackage{etoolbox}
\usepackage{subfiles}

\begin{document}

\setstretch{1.5}

\title{ Rigorous Bound on the Violation of Dynamic Reciprocity Induced by Four-Wave Mixing
}
\baselineskip=12pt
\author{Alexander D.~White}
\affiliation{E. L. Ginzton Laboratory Stanford University Stanford CA 94305 USA.}
\author{Rahul Trivedi}
\affiliation{Electrical and Computer Engineering University of Washington Seattle WA -  98195 USA}
\affiliation{Max Planck Institute of Quantum Optics Hans Kopfermann Str. 1 Garching bei Muenchen 85748}

\baselineskip=24pt

\baselineskip=24pt

\renewcommand{\baselinestretch}{1.0}
\setstretch{1}
\begin{abstract}
    Dynamic reciprocity imposes stringent performance constraints on nonlinear optical devices such as isolators and circulators. The seminal result by Shi et al. \cite{shi2015limitations} establishes that nonlinear optical devices relying on the intensity-dependent refractive index obey dynamic reciprocity for small signals with spectrally distinct fields. However, it has also been recognized that it is possible to violate dynamic reciprocity by exploiting frequency mixing processes. In this paper, we establish a rigorous upper bound on this violation that is independent of device geometry. We demonstrate that this bound captures the parameter scalings of realizable physical systems, and that under some conditions dynamic reciprocity violation can grow unbounded to achieve arbitrary nonlinear isolation. These results provide an analytically robust version of dynamic reciprocity, as well as theoretical guidance for the development of power efficient nonlinear optical isolators and circulators.
    

\end{abstract}

\maketitle



Optical isolators and circulators are critical components in laser systems, allowing for the propagation of light in one direction while prohibiting it in the other. To achieve isolation, devices must violate Lorentz reciprocity \cite{jalas2013and}. Traditionally, this has been done by breaking time reversal symmetry using magneto-optic materials \cite{turner1981fiber, du2018monolithic} or direct time-modulation of devices \cite{yu2009complete, herrmann2022mirror}. Nonlinearity offers another path to alleviating Lorentz reciprocity \cite{del2018microresonator, white2023integrated}, but as shown by Shi et al \cite{shi2015limitations}, the ways in which this can be done are quite limiting. While reciprocity can be broken with large signals, there is a small signal dynamic reciprocity that still holds.

\begin{figure}[h!]
\centering\includegraphics[width=0.95\linewidth]{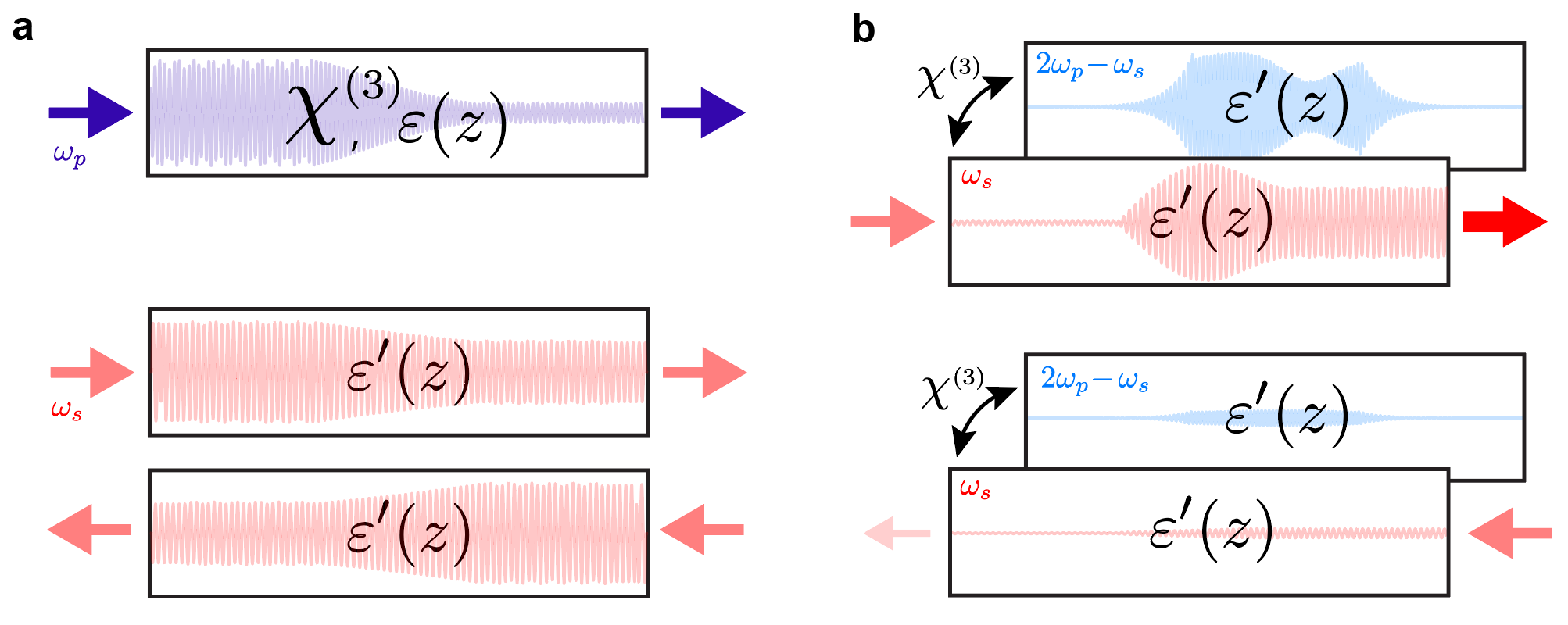}
\caption{{\bf Schematic.} {\bf a}. Illustration of dynamic reciprocity. A pump field at $\omega_p$ impinges on a device with $\chi^{(3)}$ nonlinearity and permittivity distribution $\varepsilon(z)$. For an incoming small signal $\omega_s \neq \omega_p$, the refractive index shift due to the nonlinearity can be absorbed into a modified permitivity distribution $\varepsilon'(z)$. The resulting linear time-invarient system is reciprocal. {\bf b}. Illustration of the same system when frequency mixing terms are accounted for. Small signal behavior now couples the field at $\omega_s$ to that of $2\omega_p-\omega_s$ in a direction dependant fashion, allowing for the violation of dynamic reciprocity.  }
\label{fig:Fig1}
\end{figure}

To see this, Shi et al. consider the behavior of a Kerr nonlinear device with permittivity $\varepsilon(z)$ under the influence of a large (pump) and small (signal) optical excitation, Fig 1a. If the pump and signal are spectrally distinct, the Kerr nonlinearity contributes an polarization density of $P^{NL}(\omega_s) \cong 6\epsilon_0 \chi^{(3)}(z)|E_{\omega_p}(z)|^2 E_{\omega_s}(z)$ at the signal frequency. This can be absorbed into the permittivity distribution $\varepsilon'(z) = \varepsilon(z) + 6\epsilon_0 \chi^{(3)}(z)|E_{\omega_p}(z)|^2$, leading to a reciprocal transmission of the signal through a modified refractive index profile. This means that even if a device allows for the unidirectional transmission of large excitations, small signal noise can be unintentionally transmitted backwards.

Recent experimental demonstrations of nonlinear optical isolators \cite{del2018microresonator, white2023integrated} get around dynamic reciprocity by ensuring that the reciprocal small signal transmission at frequencies around the pump is nearly zero, while the large-signal transmission of the pump itself is large. It was recognized in Ref.~\cite{shi2015limitations} that with a small signal at the same frequency as that of the pump, or with a spectral overlap with the pump, dynamic reciprocity does not apply. However, while this allows for the efficient isolation of a continuous wave pump laser, the operation is limited to a single frequency. It has also been demonstrated that frequency conversion can be used to circumvent dynamic reciprocity in larger frequency bands \cite{hua2016demonstration, de2022optical}. While a separation has been drawn in the literature between nonlinear refractive index and non-degenerate four wave mixing \cite{fan2018nonreciprocal}, these processes are inextricably linked, leading to a general violation of dynamic reciprocity. However, it remains unclear if there are any fundamental limits to the dynamic non-reciprocity induced by a wave mixing process, and how these limits can be quantified. Building upon techniques developed by \cite{kuang2020maximal}, we derive a rigorous upper bound on this violation that is independent of device geometry. We also construct a coupled mode device which can maximally violate dynamic reciprocity and achieve arbitrary isolation ratio, and asymptotically saturates the bound.

To isolate an arbitrary signal, the small signal response must also be nonreciprocal, and thus dynamic reciprocity must be violated. 
If we consider the intensity dependant refractive index contribution from the Kerr nonlinearity we have that the nonlinear polarization is $P^{NL} = 3\epsilon_0 \chi^{(3)}|\textbf{E}|^2 \textbf{E}$. Allowing the field to have components at the pump frequency $\omega_p$, signal frequency $\omega$, and converted frequency $2\omega_p - \omega$, the small signal dynamics are governed by
\begin{align}
    &\nabla^2 \textbf{E}_\omega + \omega^2 \varepsilon(\textbf{x})\textbf{E}_\omega = 2\zeta(\textbf{x})|\textbf{E}_{\omega_p}|^2 \textbf{E}_\omega +  \zeta(\textbf{x})\textbf{E}_{\omega_p}^2 \textbf{E}_{2\omega_p - \omega}^* \\
    &\nabla^2 \textbf{E}_{2\omega_p - \omega} + (2\omega_p - \omega)^2 \varepsilon(\textbf{x})\textbf{E}_{2\omega_p - \omega} = 2\zeta(\textbf{x})|\textbf{E}_{\omega_p}|^2 \textbf{E}_{2\omega_p - \omega} +  \zeta(\textbf{x})\textbf{E}_{\omega_p}^2 \textbf{E}_{\omega}^*
\end{align}
where $\zeta(\textbf{x}) = 3\epsilon_0 \chi^{(3)}(\textbf{x})$.

If $\textbf{E}_{2\omega_p - \omega} = 0$ everywhere, we revert to the model where dynamic reciprocity was originally derived \cite{shi2015limitations}. However, even if the input and output of the system at frequency $2\omega_p - \omega$ is 0, there can be a nonzero field at this frequency in the interior of the device.
In fact \emph{any} Kerr device pumped with $\omega_p$ and $\omega$ will have a nonzero field at $2\omega_p - \omega$: if $\textbf{E}_{2\omega_p - \omega} = 0$, equation (7) becomes $\zeta(\textbf{x})\textbf{E}_{\omega_p}^2 \textbf{E}^*_{\omega} = 0$, a contradiction. 
This nonzero field removes the restraint of dynamic reciprocity, as the frequency mixing term $\zeta(\textbf{x})\textbf{E}_{\omega_p}^2 \textbf{E}_{2\omega_p - \omega}^*$ is dependant on the direction of propagation, Fig 1b.
For small enough nonlinearity and pump power, this contribution is small, and we can formulate a robust version of dynamic reciprocity as a bound on the reciprocity violation. For larger nonlinearities and power, the field at $2\omega_p - \omega$ cannot be ignored and the reciprocity violation is large enough to construct isolators. 



\begin{figure}[h!]
\centering\includegraphics[width=0.95\linewidth]{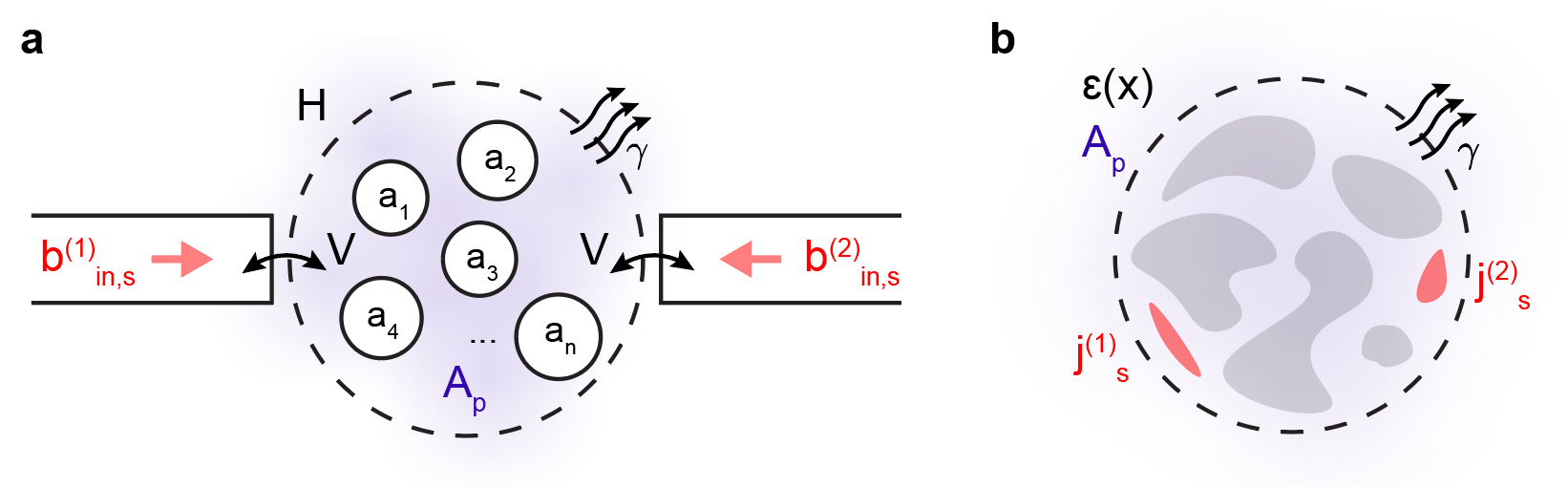}
\caption{{\bf Bound Setup.} {\bf a}. Coupled mode system: Two waveguide modes $b^{(1)}, b^{(2)}$ couple to a collection of coupled cavities $a_1 ... a_n$ with coupling matrix $V$. The cavities are governed by a hamiltonian $H$ and have an intrinsic loss rate $\gamma$. A strong pump field $A_p$ introduces nonlinearity to the system. {\bf b}. General system: Two current sources $j^{(1)}, j^{(2)}$ couple to a permittivity distribution $\varepsilon(x)$ with an intinsic loss rate $\gamma$. A strong pump field $A_p$ in the system introduces nonlinearity.}
\label{fig:Fig1}
\end{figure}

To quantify the violation of dynamic reciprocity, we consider the small signal version of lorentz reciprocity
\begin{equation}
    \Delta_{\text{DR}} = \left|\int j_1 e_2 d\textbf{x} - \int j_2 e_1 d\textbf{x} \right|.
\end{equation}
We first calculate this bound in the coupled mode framework, Fig 2a. In coupled mode theory, 
\begin{align}
    \Delta_{\text{DR}} = |s_{12} - s_{21}| = \left | b_{\text{in},s}^{(1)T}Sb_{\text{in},s}^{(2)} - b_{\text{in},s}^{(2)T}Sb_{\text{in},s}^{(1)} \right |.
\end{align}
We consider the general coupled mode system governed by
\begin{align}
    \frac{d}{dt} a(t) = -iH_{\text{eff}}a - i\chi\sum_{i,j,k}C_{i,j,k}a_i a_j a_k^* - iVb_{\text{in}}(t)
\end{align}
where $C_{i,k,j}$ is a vector describing the nonlinear overlap contributions corresponding to the interactions of modes $a_i, a_j, a_k$, and 
\begin{align}
    H_{\text{eff}} = H - \frac{i}{2}V^\dagger V- \frac{i \gamma}{2} I
\end{align}
where $H$ is the Hamiltonian governing the linear system dynamics, $V$ is the coupling matrix to the input and output waveguides, and $\gamma$ is the intrinsic loss rate.
To bound $\Delta_{\text{DR}}$ independent of $H$, we build upon the analysis of \cite{kuang2020maximal} to bound the norm of $a$, from which we obtain a bound on $\Delta_{\text{DR}}$.

For the coupled mode dynamics we find that
\begin{align}
    \Delta_{\text{DR}} \leq \frac{8\chi^2}{\gamma^2} \left\Vert V^{\text{T}} b_{\text{in,s}}^{(1)} \right\Vert \left\Vert V^{\text{T}} b_{\text{in,s}}^{(2)} \right\Vert \frac{\left\Vert M(A^2) \right\Vert^2 }{\frac{\gamma}{2} - \frac{2\chi^2}{\gamma} \left\Vert M(A^2) \right\Vert^2}, 
\end{align}
where $M$ is a matrix that depends on $A^2$ that takes into account the frequency
mixing components induced by the pump amplitude $A$.

To derive a bound that is independent of device geometry, we take a similar approach now starting from Maxwell's equations. We consider a system with an arbitrary distribution of nonlinear material, Fig 2b. We find that
\begin{align}
    \Delta_{\text{DR}} \leq 2 \alpha \beta \frac{\omega_s^2}{c^2} \left\Vert \phi_{1, \text{inc}, s} \right\Vert \left\Vert \phi_{2, \text{inc}, s} \right\Vert (2\chi |\mathcal{E}_p|^2 + \gamma \chi^2 |\mathcal{E}_p|^4)
\end{align}
where
\begin{align*}
    \alpha &= \frac{1}{|\varepsilon| Im(\varepsilon^{-1})} \\
    \beta &= \frac{\alpha}{1 - \alpha \frac{\omega_s^2}{c^2} \chi |\mathcal{E}_p|^2 [2+ \gamma\chi |\mathcal{E}_p|^2]\left\Vert G_s \right\Vert  } \\
    \gamma &= \frac{\alpha \frac{\omega_i^2}{c^2} \left\Vert G_i \right\Vert  }{1 - \alpha \frac{\omega_i^2}{c^2} 2\chi |\mathcal{E}_p|^2\left\Vert G_i \right\Vert }, 
\end{align*}
where $\epsilon$ is the permittivity of the nonlinear material, $\left\Vert\phi_{inc}\right\Vert$ is the overlap of the incident electric field with the nonlinear material, $\mathcal{E}_p$ is the maximum electric field of the of the pump inside the device, and $G$ is the Green's function of free space.

Is there a device that can maximally violate dynamically reciprocity and saturate the bounds in Eq. 7 and 8? 
Similar to \cite{hua2016demonstration} and \cite{de2022optical}, we consider a coupled mode theory of a nonlinear ring resonator with modes near the signal, pump, and converted frequency, Fig 3a. We introduce an anomalous dispersion to the ring to allow for a phase matching condition, and couple the ring to a waveguide at rate $\kappa$ at $\omega_s + 2\Delta$ and $\omega_p + \Delta$. This could be done in practice by coupling to the ring with a waveguide that cannot support a mode at $2\omega_p-\omega_s$. We also impose a loss rate $\gamma$, which prevents optical parametric oscillation at mode $2\omega_p-\omega_s + 2\Delta$ and allows us to critically couple to the other modes.

\begin{figure}[h!]
\centering\includegraphics[width=0.95\linewidth]{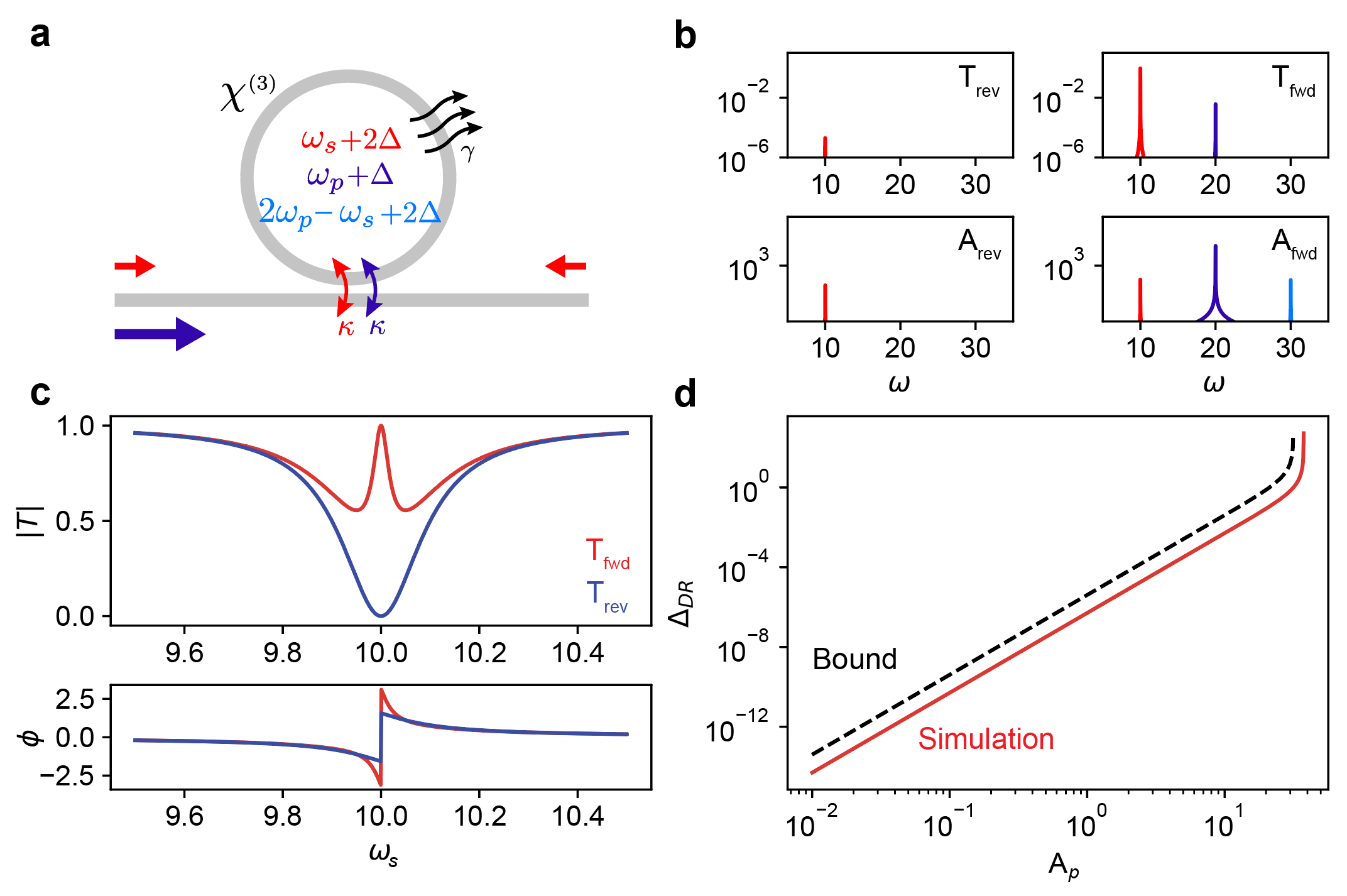}
\caption{{\bf Example with Coupled Mode Theory.} {\bf a}. Schematic of device. We couple a ring resonator with a $\chi^{(3)}$ nonlinearity with degenerate clockwise and counterclockwise modes at $\omega_s + 2\Delta, \omega_p+\Delta, 2\omega_p - \omega_s + 2\Delta$ to a waveguide. The modes near the signal $\omega_s + 2\Delta$ and pump  $\omega_p+\Delta$ frequencies are coupled to the waveguide with rate $\kappa$, and a loss rate of $\gamma$ is imposed to the frequency converted mode at $2\omega_p - \omega_s + 2\Delta$. {\bf b}. Power spectrum from nonlineared coupled mode equations. Top row shows a representative transmitted power spectrum in the reverse (left) and forward (right) direction. Bottom row shows the corresponding power spectrum in each mode. {\bf c}. Small signal transmission amplitude and phase of $\omega_s$ in the reverse and forward directions. {\bf d}. Reciprocity violation $\Delta_{\text{DR}}$ bound and simulation on resonance for varying pump power.}
\label{fig:Fig1}
\end{figure}

We send a strong pump through the waveguide at frequency $\omega_p$, and monitor the transmission of a small signal at $\omega_s$ in the forwards and backwards direction. Figure 3b shows an example of this simulated using the full nonlinear coupled mode equations. When the pump and signal are traveling in opposite directions, the frequency mixing process is not phase matched, and the signal is only affected by the cross phase modulation of the pump, leading to critical coupling to the ring and near 0 transmission. When the pump and signal are propagating in the same direction, the frequency mixing terms are phase matched and the signal is amplified, allowing for a transmission near 1. 

If we assume the signal and converted frequency are small compared to the pump, we can also derive linear coupled mode equations with a closed form solution. Fig 3c shows the transmission spectrum of this small signal model for the forwards and backwards directions at a given pump power. By setting $\kappa = \gamma$ and pumping the ring below threshold, we can achieve unity transmission with an arbitrary isolation. We note that the pump amplitude needed for the experimentally demonstrated isolators \cite{del2018microresonator, white2023integrated} to reach 3dB isolation threshold is $A \approx (\frac{\gamma}{2g})^{\frac{1}{2}}$, so these parameters are well within experimental realizability. This isolator can achieve significantly higher isolation with comprable power, and unlike in \cite{del2018microresonator, white2023integrated}, can operate outside of the large signal CW regime.

Evaluating our bound for this system, Fig 3d,
\begin{align}
    \Delta_{\text{DR}} \leq  4  \frac{4g^2 |A|^4 / \gamma^2}{1 - 4g^2 |A|^4  / \gamma^2},
\end{align}
we find it scales identically to the actual behavior of the system with all parameters and differs only by constants
\begin{align}
  \Delta_{\text{DR}} = \frac{2g^2|A|^4/\gamma^2}{1 - 2g^2|A|^4/\gamma^2}.
\end{align}
As the bound captures all of the system scalings, it is likely reflective of the true bound. Additionally, as it tracks closely to our rigorous bound, the system presented is likely a near optimal violator of dynamic reciprocity. 

While dynamic reciprocity has been extremely useful for guiding the design of nonlinear isolators, we have shown here that it can break down when inherent frequency mixing processes are accounted for. By taking these processes into account, we have derived a robust version of dynamic reciprocity formulated as a bound on the violation of Lorentz reciprocity. In cases where this violation is large, we can take advantage of this to build a new class of nonlinear optical isolators, which we have demonstrated numerically in the example presented.

\clearpage

\bibliography{References}
\clearpage


\title{Supplemental Information}

\maketitle

\section*{Coupled Mode Bound}
Here we will bound the violation of the reciprocity
\begin{align}
    \Delta_{\text{DR}} = |s_{12} - s_{21}|
\end{align}
in a general coupled mode system with third order nonlinearity. Equivalently,
\begin{align}
    \Delta_{\text{DR}} = \left | b_{\text{in},s}^{(1)\text{T}}Sb_{\text{in},s}^{(2)} - b_{\text{in},s}^{(2)\text{T}}Sb_{\text{in},s}^{(1)} \right |.
\end{align}
To do this, we will first derive a bound the signal and idler fields in the device. If there were no idler field, $\Delta_{\text{DR}}$ would be 0 as in \cite{shi2015limitations}. However, in the general case this is not true, and we will use our bound on the idler field to construct a bound on $\Delta_{\text{DR}}$.
We consider the general coupled mode system governed by
\begin{align}
    \frac{d}{dt} a(t) = -iH_{\text{eff}}a - i\chi\sum_{i,j,k}C_{i,j,k}a_i a_j a_k^* - iVb_{\text{in}}(t)
\end{align}
where $C_{i,k,j}$ is a vector describing the nonlinear overlap contributions corresponding to the interactions of modes $a_i, a_j, a_k$, and 
\begin{align}
    H_{\text{eff}} = H - \frac{i}{2}V^\dagger V- \frac{i \gamma}{2} I
\end{align}
Let us take 
\begin{align}
    a(t) = Ae^{-i\omega_p t} + a_s e^{-i\omega_s t} + a_i e^{-i\omega_i t}
\end{align}
where $\omega_i = 2\omega_p - \omega_s$.

Let $A \gg a_s, a_i$, then the coupled mode equations become
\begin{subequations}
\begin{align}
    -i\omega_s a_s = -iH_{\text{eff}}a_s - 2i\chi\text{diag}(|XA|^2)a_s - iVb_{\text{in},s} - i\chi M(A^2)a_i^* \\
     \label{aieqn}
     -i\omega_i a_i = -iH_{\text{eff}}a_i - 2i\chi\text{diag}(|XA|^2)a_i - iVb_{\text{in},s} - i\chi M(A^2)a_s^*
\end{align}
\end{subequations}
where $X$ is a matrix corresponding to the magnitude of cross phase modulation effect of the pump $A$, and $M$ is a matrix that depends on $A^2$ that takes into account the frequency mixing components induced by $A$.
To get a bound on the norm of the field, we can multiple both sides of the first equation by $a_s^\dagger$ and take the real part to get
\begin{subequations}
\begin{align}
    0 = \text{Re}\left[ -ia_s^\dagger H_{\text{eff}}a_s \right] - \text{Re}\left[ -ia_s^\dagger Vb_{\text{in},s} \right] - \text{Re}\left[ -ia_s^\dagger \chi M(A^2)a_i^* \right]
\end{align}
\end{subequations}
As
\begin{align}
    \text{Re}\left[ ia_s^\dagger H_{\text{eff}}a_s \right] = \frac{1}{2} \text{Re}\left[a_s^\dagger V^\dagger V a_s \right] + \frac{\gamma}{2} \left\Vert a_s \right\Vert^2 \\
    \text{Re}\left[a_s^\dagger V^\dagger V a_s \right] \geq 0
\end{align}
we have that
\begin{align}
    \frac{\gamma}{2} \left\Vert a_s \right\Vert^2 \leq \text{Re}\left[ -ia_s^\dagger Vb_{\text{in},s} \right] + \text{Re}\left[ -ia_s^\dagger \chi M(A^2)a_i^* \right]
\end{align}
rearranging the last term we get
Which with Cauchy Schwartz we can bound this with
\begin{align}
    \frac{\gamma}{2} \left\Vert a_s \right\Vert^2 \leq \left\Vert a_s \right\Vert \left\Vert Vb_{\text{in},s} \right\Vert  + \chi \left\Vert a_s \right\Vert \left\Vert a_i \right\Vert \left\Vert M(A^2) \right\Vert.
\end{align}
Repeating this analysis with equation \ref{aieqn} and $a_i^\dagger$ we find
\begin{align}
    \frac{\gamma}{2} \left\Vert a_i \right\Vert^2 \leq \chi \left\Vert a_s \right\Vert \left\Vert a_i \right\Vert \left\Vert M(A^2) \right\Vert.
\end{align}
Combining these bounds we find that
\begin{subequations}
\begin{align}
    \left\Vert a_s \right\Vert \leq \frac{\left\Vert V b_{\text{in,s}} \right\Vert}{\frac{\gamma}{2} - \frac{2\chi^2}{\gamma} \left\Vert M(A^2) \right\Vert^2} \\
    \left\Vert a_i \right\Vert \leq \frac{\frac{2\chi^2}{\gamma} \left\Vert M(A^2) \right\Vert \left\Vert V b_{\text{in,s}} \right\Vert}{\frac{\gamma}{2} - \frac{2\chi^2}{\gamma} \left\Vert M(A^2) \right\Vert^2}
\end{align}
\end{subequations}
We note that these bounds hold in the weak nonlinearity limit, i.e. when the denominator is positive. 

With these bounds on the fields inside the device, we can begin constructing a bound on $\Delta_{\text{DR}}$. We start by considering the output of the device,
\begin{align}
    b_{\text{out}, s} = S b_{\text{in},s} - iVa_s.
\end{align}
We define a new function that absorbs the frequency preserving portion of the nonlinearity into $H$
\begin{align}
    \Tilde{H}_{\text{eff}}(A) = H_{\text{eff}} - 2i\chi \text{diag}(|CA|^2).
\end{align}
Our coupled mode equation for $a_s$ then becomes
\begin{align}
    -i\omega_s a_s = -i\Tilde{H}_{\text{eff}} a_s - iV b_{\text{in,s}} -i\chi M(A^2)a_i^*
\end{align}
We can solve this equation to obtain
\begin{align}
    a_s = [i\Tilde{H}_{\text{eff}}(A) - i\omega_s I]^{-1}(-iVb_{\text{in,s}}) - \chi [i\Tilde{H}_{\text{eff}}(A) - i\omega_s I]^{-1}M(A^2)a_i^*
\end{align}
Then $b_{\text{out}}$ becomes
\begin{align}
    b_{\text{out}} = \Tilde{S}(A)b_{\text{in,s}} + \chi V[i\Tilde{H}_{\text{eff}}(A) - i\omega_s I]^{-1}M(A^2)a_i^*
\end{align}
where $\Tilde{S}(A) = S - V[i\Tilde{H}_{\text{eff}}(A) - iV\omega_s I]^{-1}(-iVb_{\text{in,s}})$ is the dynamically reciprocal portion of the scattering matrix. Plugging this into our equation for reciprocity violation and applying Cauchy Schwartz we find that
\begin{align}
\begin{split}
    \Delta_{\text{DR}} \leq \chi \left\Vert (b_{\text{in}, s}^{(1)})^{\text{T}} V [i\Tilde{H}_{\text{eff}}(A) - i\omega_s I]^{-1}] \right\Vert  \left\Vert M(A^2) \right\Vert \left\Vert a_i^{(2)} \right\Vert \\
    + \chi \left\Vert (b_{\text{in}, s}^{(2)})^{\text{T}} V [i\Tilde{H}_{\text{eff}}(A) - i\omega_s I]^{-1}] \right\Vert  \left\Vert M(A^2) \right\Vert \left\Vert a_i^{(1)} \right\Vert
\end{split}
\end{align}
To bound the first term we take
\begin{align}
    \beta =  \left( \left[i\Tilde{H}_{\text{eff}}(A) - i\omega_s I\right]^{-1} \right)^{\text{T}} V^{\text{T}} b_{\text{in}, s}^{(1)}
\end{align}
Then
\begin{align}
    i\Tilde{H}_{\text{eff}}^{\text{T}}(A)\beta - i\omega_s \beta = V^{\text{T}} b_{\text{in}, s}^{(1)}
\end{align}
Multiplying both sides by $\beta^\dagger$ we get
\begin{align}
    -i\beta^\dagger\Tilde{H}_{\text{eff}}^{\text{T}}(A)\beta - i\omega_s \left\Vert \beta \right\Vert^2 = \beta^\dagger V^{\text{T}}b_{\text{in}, s}^{(1)}
\end{align}
Taking the real part we get
\begin{align}
    \frac{\gamma}{2} \left\Vert \beta \right\Vert^2 = \text{Re}\left[ \beta^* V^{\text{T}} b^{(1)}_{\text{in},s} \right]
\end{align}
and thus
\begin{align}
    \left\Vert \beta \right\Vert \leq \frac{2 \left\Vert V^{\text{T}} b^{(1)}_{\text{in},s}\right\Vert}{\gamma}
\end{align}
Plugging this and our previous bound for $a_i$, we arrive at
\begin{align}
    \Delta_{\text{DR}} \leq \frac{8\chi^2}{\gamma^2} \left\Vert V^{\text{T}} b_{\text{in,s}}^{(1)} \right\Vert \left\Vert V^{\text{T}} b_{\text{in,s}}^{(2)} \right\Vert \frac{\left\Vert M(A^2) \right\Vert^2 }{\frac{\gamma}{2} - \frac{2\chi^2}{\gamma} \left\Vert M(A^2) \right\Vert^2} 
\end{align}

\section*{Coupled Mode Theory of Ring Isolator}

The full nonlinear coupled mode equations of the ring device described in the main text are given by:
\begin{subequations}
\begin{align}
    \dot{a}_+ &= -(i\omega_a + \kappa/2 + \gamma/2) a_+ + ig(2P_{tot} - |a_+|^2) a_+ + igb_+^2 c_+^* + igc_+^2 b_+^* - i\sqrt{\kappa} s_{\omega_p,+} \\
    \dot{a}_- &= -(i\omega_a + \kappa/2 + \gamma/2) a_+ + ig(2P_{tot} - |a_-|^2) a_- + igb_-^2 c_-^* + igc_-^2 b_-^* - i\sqrt{\kappa} s_{\omega_p,-} \\
    \dot{b}_+ &= -(i\omega_b + \kappa/2 + \gamma/2) b_+ + ig(2P_{tot} - |b_+|^2) b_+ + iga_+^2 c_+^* + igc_+^2 a_+^* - i\sqrt{\kappa} s_{\omega_s,+} \\
    \dot{b}_- &= -(i\omega_b + \kappa/2 + \gamma/2) b_- + ig(2P_{tot} - |b_-|^2) b_- + iga_-^2 c_-^* + igc_-^2 a_-^* - i\sqrt{\kappa} s_{\omega_s,-} \\
    \dot{c}_+ &= -(i\omega_c + \gamma/2) c_+ + ig(2P_{tot} - |c_+|^2) c_+ + igb_+^2 a_+^* + iga_+^2 b_+^* \\
    \dot{c}_- &= -(i\omega_c + \gamma/2) c_- + ig(2P_{tot} - |c_-|^2) c_+ + igb_-^2 a_-^* + iga_-^2 b_-^*,
\end{align}
\end{subequations}

where $g = 3\epsilon_0\chi^{(3)}$, $P_{tot} = |a_+|^2 + |a_-|^2 + |b_+|^2 + |b_-|^2 + |c_+|^2 + |c_-|^2$, $\omega_a \approx \omega_p$, $\omega_b \approx \omega_s$,  $\omega_c \approx 2\omega_a - \omega_b$, . The clockwise (+) and counterclockwise (-) modes of the ring contribute cross phase modulation terms to each other, but do not contain any mixed frequency conversion terms. This is because the resulting wave would have a mismatched momentum, preventing this conversion. If we let each mode have the same momentum, $a_+ = \alpha_+ e^{-i \omega_a t+ikx}$, $b_- = \beta_- e^{-i \omega_b t -ikx}$, then $a^2_+b^*_- = \alpha_+^2 \beta_-^* e^{-i(2\omega_a-\omega_b)t+3ikx}$. While the frequency of the resultant wave is correct, the momentum is too large to couple to the mode $c$ by a factor of 3.

If we consider driving only the clockwise modes of the pump ($a$) and signal ($b$), our coupled mode equations become:
\begin{subequations}
\begin{align}
    \dot{a}_+ &= -(i\omega_a + \kappa/2 + \gamma/2) a_+ + ig(|a_+|^2 + 2|b_+|^2 + 2|c_+|^2) a_+ + igb_+^2 c_+^* + igc_+^2 b_+^* - i\sqrt{\kappa} s_{\omega_p,+} \\
    \dot{b}_+ &= -(i\omega_b + \kappa/2 + \gamma/2) b_+ + ig(2|a_+|^2 + |b_+|^2 + 2|c_+|^2) b_+ + iga_+^2 c_+^* + igc_+^2 a_+^* - i\sqrt{\kappa} s_{\omega_s,+} \\
    \dot{c}_+ &= -(i\omega_c + \gamma/2) c_+ + ig(2|a_+|^2 + 2|b_+|^2 + |c_+|^2) c_+ + igb_+^2 a_+^* + iga_+^2 b_+^*.
\end{align}
\end{subequations}

And if we drive the clockwise pump and counterclockwise signal we have:
\begin{subequations}
\begin{align}
    \dot{a}_+ &= -(i\omega_a + \kappa/2 + \gamma/2) a_+ + ig(|a_+|^2 + 2|b_-|^2) a_+ + igb_+^2 c_+^* + igc_+^2 b_+^* - i\sqrt{\kappa} s_{\omega_p,+} \\
    \dot{b}_- &= -(i\omega_b + \kappa/2 + \gamma/2) b_- + ig(2|a_+|^2 + |b_-|^2) b_-  - i\sqrt{\kappa} s_{\omega_s,-}.
\end{align}
\end{subequations}

If we let $a_+ = Ae^{-i\omega_p t}, a_- = 0$, and $|b|, |c| << |A|$, the linearized forward equations for a small signal in $b$ and $c$ become:
\begin{subequations}
\begin{align}
    \dot{b}_+ &= -(i(\omega_b - 2g|A|^2) + \kappa/2 + \gamma/2) b_+ + igA^2 e^{-i2\omega_p t} c_+^* - i\sqrt{\kappa} s_{\omega_s,+} \\
    \dot{c}_+ &= -(i(2\omega_c - 2g|A|^2) + \gamma/2) c_+ + igA^2 e^{-i2\omega_p t} b_+^*,
\end{align}
\end{subequations}
and the backwards equation becomes:
\begin{align}
    \dot{b}_- &= -(i(\omega_b - 2g|A|^2) + \kappa/2 + \gamma/2) b_- - i\sqrt{\kappa} s_{\omega_s,-}.
\end{align}

To solve for the transmission, we let $b = \beta e^{-i\omega_s t}$ and $c = \zeta e^{-i(2\omega_p - \omega_s) t}$. We can then fourier transform the above equations and find that:

\begin{subequations}
\begin{align}
    b_+(\omega) &= \frac{-i\sqrt{\kappa}s_{\omega_s,+}(\omega)}{i(\omega_b - 2g|A|^2-\omega)-\kappa/2 - \gamma/2 + \Gamma} \\
    \Gamma &= \frac{g^2 |A|^4}{\gamma/2 + i(2\omega_p - \omega - \omega_c + 2g|A|^2)},
\end{align}
\end{subequations}
and

\begin{align}
    b_-(\omega) &= \frac{-i\sqrt{\kappa}s_{\omega_s,-}(\omega)}{i(\omega_b - 2g|A|^2-\omega)-\kappa/2 - \gamma/2}.
\end{align}

The dynamics is identical up to the gain term $\Gamma$ from the momentum matched mode in the forwards direction. If $\text{Re}(\Gamma) \geq \kappa/2 + \gamma/2$, the continuous wave assumption breaks down and the amplitude of $b_+$ grows unbounded as $\sinh(t)$.

The output fields at the signal frequency are given as:
\begin{subequations}
\begin{align}
    s_{out, \omega_s,+} &= s_{\omega_s,+} - i \sqrt{\kappa}b_+ \\
    s_{out, \omega_s,-} &= s_{\omega_s,-} - i \sqrt{\kappa}b_-.
\end{align}
\end{subequations}

Thus the forwards and backwards small signal transmission are 
\begin{subequations}
\begin{align}
    \text{T}_f = \left|\frac{s_{out, \omega_s,+}}{s_{\omega_s,+}}\right|^2 &= \left|1 + \frac{\kappa}{i(\omega_b - 2g|A|^2-\omega)-\kappa/2 - \gamma/2 + \Gamma} \right|^2\\
    \text{T}_r = \left|\frac{s_{out, \omega_s,-}}{s_{\omega_s,-}}\right|^2 &= \left|1 + \frac{\kappa}{i(\omega_b - 2g|A|^2-\omega)-\kappa/2 - \gamma/2} \right|^2.
\end{align}
\end{subequations}

If we set the resonance frequencies to the ideal phase matching condition:
\begin{subequations}
\begin{align}
    \omega_a &= \omega_p + g|A|^2 \\
    \omega_b &= \omega_s + 2g|A|^2 \\
    \omega_c &= 2\omega_p - \omega_s + 2g|A|^2,
\end{align}
\end{subequations}
the transmission simplifies to 
\begin{subequations}
\begin{align}
    \text{T}_f &= \left|1 + \frac{\kappa}{i(\omega_s-\omega)-\kappa/2 - \gamma/2 + \Gamma} \right|^2,
    \Gamma = \frac{g^2 |A|^4}{\gamma/2 + i(\omega_s - \omega)} \\
    \text{T}_r &= \left|1 + \frac{\kappa}{i(\omega_s-\omega)-\kappa/2 - \gamma/2} \right|^2.
\end{align}
\end{subequations}
On resonance this simplifies further to 
\begin{subequations}
\begin{align}
   \text{T}_f &= \left|1 + \frac{\kappa}{2g^2|A|^4/\gamma - \kappa/2 - \gamma/2} \right|^2\\
    \text{T}_r &= \left|1 - \frac{\kappa}{ \kappa/2 + \gamma/2} \right|^2
\end{align}
\end{subequations}
In the critically coupled case where $\kappa = \gamma$ we get
\begin{subequations}
\begin{align}
   \text{T}_f &= \left|\frac{2g^2|A|^4/\gamma^2}{1 - 2g^2|A|^4/\gamma^2} \right|^2\\
    \text{T}_r &= 0
\end{align}
\end{subequations}
To evaluate the bound for this problem, we have that
\begin{align}
    V^{\text{T}} = \begin{pmatrix}
    \sqrt{\kappa} & 0& 0& 0\\
    0 & \sqrt{\kappa}& 0& 0
    \end{pmatrix}, b_{\text{in}}^{(1)} = 
    \begin{pmatrix}
    1 & 0
    \end{pmatrix}, b_{\text{in}}^{(2)} = 
    \begin{pmatrix}
    0 & 1
    \end{pmatrix}, M = 
    \begin{pmatrix}
    0 & A^2& 0& 0\\
    A^2 & 0& 0& 0\\
    0 & 0& 0& 0\\
    0 & 0& 0& 0
    \end{pmatrix}
\end{align}

Plugging this into our bound equation we get

\begin{align}
    \Delta_{\text{DR}} \leq \frac{8g^2}{\gamma^2} \kappa \frac{|A|^4 }{\frac{\gamma}{2} - \frac{2g^2}{\gamma} |A|^4} = 4 \frac{\kappa}{\gamma} \frac{4 \frac{g^2}{\gamma^2} |A|^4 }{1 - 4 \frac{g^2}{\gamma^2} |A|^4}
\end{align}

as we set $\kappa = \gamma$ for critical coupling, we have that

\begin{align}
    \Delta_{\text{DR}} \leq  4  \frac{4g^2 |A|^4 / \gamma^2}{1 - 4g^2 |A|^4  / \gamma^2}
\end{align}

\section*{Derivation of General Bound}

To derive the bound, we begin with the definition of Lorentz reciprocity, 

\begin{equation}
    \int j_1 e_2 d\textbf{x} = \int j_2 e_1 d\textbf{x}.
\end{equation}

We will bound the violation of this equation,  

\begin{equation}
    \left|\int_{\textbf{x} \in S_1} j_{1, \omega} e_{2, \omega} d\textbf{x} - \int_{\textbf{x} \in S_2} j_{2, \omega} e_{1, \omega} d\textbf{x}\right|.
\end{equation}.

We first start by finding bounds for the signal and idler fields inside the device.

Let the device geometry be parameterized by the function $\Theta(\textbf{x}), \Theta: \mathbb{R} \rightarrow \{0,1\} $, the complex dielectric constant $\varepsilon$, and the third order nonlinearity $\chi$ such that:
\begin{align}
    \epsilon(\textbf{x}) = \epsilon_0 + \Theta(\textbf{x})[\varepsilon + \chi|E_p(\textbf{x})|^2]
\end{align}

For a small signal source $J$ at frequency $\omega_s$ we have that:
\begin{subequations}
\begin{align}
\begin{split}
    e_s(\textbf{x}) =& \int  G_s(\textbf{x}, \textbf{x}') J(\textbf{x}')d\textbf{x}' \\
    &+ \frac{\omega_s^2}{c^2} \int  G_s(\textbf{x}, \textbf{x}') [\varepsilon + 2\chi|E_p(\textbf{x}')|^2]\Theta(\textbf{x}')e_s d\textbf{x}' \\
    &+ \frac{\omega_s^2}{c^2}\int  G_s(\textbf{x}, \textbf{x}') \chi E_p(\textbf{x}')^2\Theta(\textbf{x}')e_i^* d\textbf{x}'
\end{split}
\end{align}
\begin{align}
\begin{split}
        e_i(\textbf{x}) =& \frac{\omega_i^2}{c^2} \int  G_i(\textbf{x}, \textbf{x}') [\varepsilon + 2\chi|E_p(\textbf{x}')|^2]\Theta(\textbf{x}')e_i d\textbf{x}' \\
    &+ \frac{\omega_i^2}{c^2}\int  G_i(\textbf{x}, \textbf{x}')\chi E_p(\textbf{x}')^2 \Theta(\textbf{x}')e_s^* d\textbf{x}'
\end{split}
\end{align}
\end{subequations}
where $ G_s(\textbf{x})$ is the Green's function of free space at frequency $\omega_s$. 

We define a new function $\phi_s(\textbf{x}) = \Theta(\textbf{x}')e_i(\textbf{x})$. Multiplying both sides by $\phi_s(\textbf{x})$ and integrating we get that

\begin{align}
\begin{split}
    \int |\phi_s(\textbf{x})|^2 d\textbf{x} =&  \int \phi_s^*(\textbf{x}) e_{s,  \text{inc}} d\textbf{x} \\
    &+ \frac{\omega_s^2}{c^2} \int \int \phi_s^*(\textbf{x})  G_s(\textbf{x}, \textbf{x}') [\varepsilon + 2\chi|E_p(\textbf{x}')|^2]\phi_s(\textbf{x}') d\textbf{x}'d\textbf{x} \\
    &+ \frac{\omega_s^2}{c^2}\int \int \phi_s^*(\textbf{x})  G_s(\textbf{x}, \textbf{x}') [2\chi E_p(\textbf{x}')^2]\phi_i^* d\textbf{x}'d\textbf{x}
\end{split}
\end{align}
where $e_{s,  \text{inc}} = \int  G_s(\textbf{x}, \textbf{x}') J(\textbf{x}')d\textbf{x}'$. On the left hand side we multiply by an additional factor of $\Theta(\textbf{x})$ when multiplying $\phi(\textbf{x})$. 

Dividing by $\varepsilon$ and taking the imaginary part of both sides we get that
\begin{align}
\begin{split}
    |\varepsilon| Im(\varepsilon^{-1}) \left\Vert\phi_s(\textbf{x})\right\Vert^2 =&  Im(\int \phi_s^*(\textbf{x}) e_{s,  \text{inc}} d\textbf{x} \\
    &+ \int \int \phi_s^*(\textbf{x})  G_s(\textbf{x}, \textbf{x}') [2\chi|E_p(\textbf{x}')|^2]\phi_s(\textbf{x}') d\textbf{x}'d\textbf{x} \\
    &+ \frac{\omega_s^2}{c^2}\int\int \phi_s^*(\textbf{x})  G_s(\textbf{x}, \textbf{x}') [2\chi E_p(\textbf{x}')^2]\phi_i^* d\textbf{x}'d\textbf{x}).
\end{split}
\end{align}
where on the right hand side the term $Im(\int \phi_s^*(\textbf{x}) * G(\textbf{x},\textbf{x}') \phi_s(\textbf{x}')d\textbf{x}'d\textbf{x})= 0$. We can then use the Cauchy Schwartz inequality to get that
\begin{align}
    |\varepsilon| Im(\varepsilon^{-1}) \left\Vert \phi_s(\textbf{x}) \right\Vert^2 \leq& \left\Vert\phi_{s,  \text{inc}}\right\Vert  \left\Vert\phi_s\right\Vert + \frac{\omega_s^2}{c^2} 2\chi |\mathcal{E}_p|^2 \left\Vert  G_s\right\Vert  \left\Vert\phi_s\right\Vert^2 + \frac{\omega_s^2}{c^2} \chi |\mathcal{E}_p|^2 \left\Vert  G_s\right\Vert  \left\Vert\phi_s\right\Vert  \left\Vert\phi_i\right\Vert.
\end{align}
where $\mathcal{E}_p$ is the maximum value of $E_p(\textbf{x})$. Following the same procedure with the idler field we get that
\begin{align}
    |\varepsilon| Im(\varepsilon^{-1})\left\Vert\phi_i(\textbf{x}) \right\Vert ^2 \leq&  \frac{\omega_s^2}{c^2} 2\chi |\mathcal{E}_p|^2\left\Vert  G_i \right\Vert   \left\Vert\phi_i \right\Vert ^2 + \frac{\omega_s^2}{c^2} \chi |\mathcal{E}_p|^2\left\Vert  G_i \right\Vert   \left\Vert\phi_i \right\Vert   \left\Vert\phi_s \right\Vert .
\end{align}
It follows that
\begin{align}
   \left\Vert\phi_i \right\Vert  \leq \frac{\alpha \frac{\omega_i^2}{c^2} \chi |\mathcal{E}_p|^2\left\Vert G_i \right\Vert  }{1 - \alpha \frac{\omega_i^2}{c^2} 2\chi |\mathcal{E}_p|^2\left\Vert G_i \right\Vert  }\left\Vert\phi_s \right\Vert
\end{align}
where $\alpha = \frac{1}{|\varepsilon| Im(\varepsilon^{-1})}$, as long as the denominator is positive. Plugging this into the expression for $ \left\Vert \phi_s \right\Vert $, we find
\begin{subequations}
\begin{align}
\label{fieldBounds}
   \left\Vert\phi_s \right\Vert  &\leq \frac{\alpha\left\Vert\phi_{s, \text{inc}} \right\Vert  }{1 - \alpha \frac{\omega_s^2}{c^2} \chi |\mathcal{E}_p|^2 [2+ \gamma\chi |\mathcal{E}_p|^2]\left\Vert G_s \right\Vert  } \\
   \left\Vert\phi_i \right\Vert  &\leq \frac{\gamma \chi |\mathcal{E}_p|^2 \alpha\left\Vert\phi_{s, \text{inc}} \right\Vert  }{1 - \alpha \frac{\omega_s^2}{c^2} \chi |\mathcal{E}_p|^2 [2+ \gamma\chi |\mathcal{E}_p|^2]\left\Vert G_s \right\Vert  }
\end{align}
\end{subequations}
where 
\begin{align}
    \gamma = \frac{\alpha \frac{\omega_i^2}{c^2} \left\Vert G_i \right\Vert  }{1 - \alpha \frac{\omega_i^2}{c^2} 2\chi |\mathcal{E}_p|^2\left\Vert G_i \right\Vert  }
\end{align}
With these bounds in hand, we can return to the expression
\begin{align}
    \Delta_{\text{DR}} = \left|\int_{\textbf{x} \in S_1} j_{1, s} e_{2, s} d\textbf{x} - \int_{\textbf{x} \in S_2} j_{2, s} e_{1, s} d\textbf{x}\right|.
\end{align}.
With the Green's function formalism
\begin{align}
\begin{split}
    \int j_{1,s}(\textbf{x})e_{2, s}(\textbf{x}) d\textbf{x} =& \int j_{1,s}(\textbf{x}) G_{s, \text{lin}}(\textbf{x}, \textbf{x}') j_{2,s}(\textbf{x}')d\textbf{x}'d\textbf{x} \\
    &+ \frac{\omega_s^2}{c^2} \int \int j_{1,s}(\textbf{x}) G_{s, \text{lin}}(\textbf{x}, \textbf{x}') 2\chi|E_p(\textbf{x}')|^2 \phi_{2,s}(\textbf{x}') d\textbf{x}' d\textbf{x} \\
    &+ \frac{\omega_s^2}{c^2}\int \int j_{1,s}(\textbf{x}) G_{s, \text{lin}}(\textbf{x}, \textbf{x}') \chi E_p(\textbf{x}')^2\phi_{2,i}^*(\textbf{x}') d\textbf{x}' d\textbf{x}
\end{split}
\end{align}
where $G_{s, \text{lin}}$ is the Green's function of the linear dielectric contribution to the device. Note that the first term cancels in our expression for $\Delta_{\text{DR}}$.

Integrating over x we have that 
\begin{equation}
    \int \int j_{1,s}(\textbf{x}) G_{s, \text{lin}}(\textbf{x}, \textbf{x}') \Theta(\textbf{x}') f(\textbf{x}') d\textbf{x}' d\textbf{x} = \int \Theta(\textbf{x}') e_{1,\text{lin}, s}(\textbf{x}') \Theta(\textbf{x}') f(\textbf{x}') d\textbf{x}'.
\end{equation}
where $e_{1,\text{lin}, s}$ is the electric field of the device with $\chi = 0$.
Thus we have that
\begin{align}
\begin{split}
    \Delta_{\text{DR}} \leq& \frac{\omega_s^2}{c^2} \left\Vert \phi_{1, \text{lin}, s} \right\Vert (|\varepsilon| + 2\chi |\mathcal{E}_p|^2) \left\Vert \phi_{2, s} \right\Vert + \frac{\omega_s^2}{c^2} \left\Vert \phi_{1, \text{lin}, s} \right\Vert \chi |\mathcal{E}_p|^2 \left\Vert \phi_{2, i} \right\Vert \\
    &+  \frac{\omega_s^2}{c^2} \left\Vert \phi_{2, \text{lin}, s} \right\Vert (|\varepsilon| + 2\chi |\mathcal{E}_p|^2) \left\Vert \phi_{1, s} \right\Vert + \frac{\omega_s^2}{c^2} \left\Vert \phi_{2, \text{lin}, s} \right\Vert \chi |\mathcal{E}_p|^2 \left\Vert \phi_{1, i} \right\Vert
\end{split}
\end{align}
From equation \ref{fieldBounds}, we can set $\chi$ to 0 and get a field bound
\begin{equation}
    \left\Vert \phi_{\text{lin}, s} \right\Vert \leq \alpha  \left\Vert\phi_{\text{inc}, s} \right\Vert
\end{equation}
Plugging in our field bounds, we get that
\begin{align}
    \Delta_{\text{DR}} \leq 2 \alpha \beta \frac{\omega_s^2}{c^2} \left\Vert \phi_{1, \text{inc}, s} \right\Vert \left\Vert \phi_{2, \text{inc}, s} \right\Vert (2\chi |\mathcal{E}_p|^2 + \gamma \chi^2 |\mathcal{E}_p|^4)
\end{align}
where
\begin{align*}
    \alpha &= \frac{1}{|\varepsilon| Im(\varepsilon^{-1})} \\
    \beta &= \frac{\alpha}{1 - \alpha \frac{\omega_s^2}{c^2} \chi |\mathcal{E}_p|^2 [2+ \gamma\chi |\mathcal{E}_p|^2]\left\Vert G_s \right\Vert  } \\
    \gamma &= \frac{\alpha \frac{\omega_i^2}{c^2} \left\Vert G_i \right\Vert  }{1 - \alpha \frac{\omega_i^2}{c^2} 2\chi |\mathcal{E}_p|^2\left\Vert G_i \right\Vert }. 
\end{align*}

\end{document}